\documentclass{article}

\usepackage{natbib}
\usepackage{graphicx}
\usepackage{multicol,multirow}
\usepackage{amsmath,amssymb,amsfonts,amsthm,amscd}
\usepackage[utf8]{inputenc}
\usepackage{textcomp}
\usepackage{xcolor}
\usepackage[colorlinks,allcolors=blue]{hyperref}

\newtheorem{theorem}{Theorem}[section]

\newtheorem{corollary}[theorem]{Corollary}
\newtheorem{proposition}[theorem]{Proposition}
\theoremstyle{definition}

\newtheorem{definition}[theorem]{Definition}
\numberwithin{equation}{section}

\usepackage{enumitem}
\usepackage{placeins}
\usepackage[margin=1in]{geometry}
\usepackage{bbm}
\newcommand{\E}[2]{\text{E}_{#1}\left[#2\right]}
\newcommand{\EE}[1]{\text{E}\left[#1\right]}
\newcommand{\keywords}[1]{\par\noindent\textbf{Keywords: }#1}

\begin{document}

\title{Adjusted Similarity Measures and a Violation of Expectations}
\date{\vspace{-5ex}}
\author{William L. Lippitt and Edward J. Bedrick and Nichole E. Carlson}
\maketitle

\abstract{Adjusted similarity measures, such as Cohen's kappa for inter-rater reliability and the adjusted Rand index used to compare clustering algorithms, are a vital tool for comparing discrete labellings. These measures are intended to have the property of 0 expectation under a null distribution and maximum value 1 under maximal similarity to aid in interpretation. Measures are frequently adjusted with respect to the permutation distribution for historic and analytic reasons. There is currently renewed interest in considering other null models more appropriate for context, such as clustering ensembles permitting a random number of identified clusters. The purpose of this work is two---fold: (1) to generalize the study of the adjustment operator to general null models and to a more general procedure which includes statistical standardization as a special case and (2) to identify sufficient conditions for the adjustment operator to produce the intended properties, where sufficient conditions are related to whether and how observed data are incorporated into null distributions. We demonstrate how violations of the sufficient conditions may lead to substantial breakdown, such as by producing a non-positive measure under traditional adjustment rather than one with mean 0, or by producing a measure which is deterministically 0 under statistical standardization.%200/200 words
}


\section{Introduction}\label{sec:intro}

Suppose we have two categorical variables we wish to compare, such as diagnostic assessments using two different assessment techniques or a ground truth labelling versus labels generated through unsupervised clustering or predictive modelling. We might use an index, that is a function of the joint empirical distribution of the two variables, to quantify their similarity. A plethora of indices have been proposed in the literature for different contexts, such as Cohen's 
$\kappa$
 \citep{cohen1960coefficient} assessing agreement when variables have the same set of possible values, variations of the Rand index \citep{rand1971objective,hubert1985comparing} and Mutual Information \citep{shannon1948mathematical,romano2016adjusting,romano2014standardized} when variables assign cluster membership, and Goodman and Kruskal's $\gamma$ \citep{goodman1979measures} measuring rank correlation when variables are ordinal. Efforts have been made to understand relationships between indices and their properties \citep{albatineh2006similarity,meilǎ2005comparing,warrens2013association,brusco2021comparison,arinik2021characterizing} in support of identifying redundancies and unwanted behavior as well as matching applied contexts with appropriate indices.

Of long-standing interest is the adjustment of indices for `random chance.' This is accomplished through subtraction of an expected value with respect to a null model used to describe expected similarity due to `random chance' followed by normalization by a maximum \citep{scott1955reliability,cohen1960coefficient,hubert1985comparing,krippendorff1987,
davenport1991phi,albatineh2006similarity,vinh2009information,warrens2013association}. Adjusted indices are thus designed to take values near 0 when compared variables are unassociated and maximal, optimal value 1, properties which support ease of interpretation and comparison \citep{vinh2009information,albatineh2006similarity,warrens2013association}. Adjustment thus requires specification of a null model and a normalization. 

It is standard to consider the permutation model, also known as the hypergeometric model, as the null model for random chance \citep{hubert1985comparing,albatineh2006similarity,vinh2009information,romano2016adjusting}. Other models have been considered for specific contexts \citep{jeub2018multiresolution,sundqvist2023adjusting,cohen1960coefficient,
krippendorff1987,scott1955reliability,li2024estimating}, such as multinomial distributions featuring independence of variables and clustering ensembles allowing variation in the number of clusters. More broadly, \cite{krippendorff1987} elucidates how different modelling assumptions lead to substantially different interpretations for what an index measures and \cite{gates2017impact} show the null model used in adjustment can have a substantial impact on conclusions. As such, null models must be chosen carefully for context.

In this work, we generalize foundational studies of the properties of the operation of adjustment for chance \citep{albatineh2006similarity,warrens2013association} which traditionally focused on the permutation and related models and specific normalizations. We consider a generalized adjustment with respect to any null model and valid choice of normalization. Generalized adjustment thus also covers statistical standardization, recently considered by \cite{romano2014standardized,romano2016adjusting} for clustering, as a special case.

Through generalized study of adjustment, we contribute the following:
\begin{itemize}[noitemsep,topsep=0pt]
	\item sufficient conditions for the adjustment operator to exhibit desired properties, such as:
	\begin{itemize}[noitemsep,topsep=0pt]
		\item adjusted indices having mean 0 under the null distribution,
		\item standardized indices also having variance 1 under the null distribution, and
		\item the adjustment operator being idempotent (no need for repeated adjustment);
	\end{itemize}
	 \item characterization of sufficient conditions in terms of whether and how the null distribution is data-driven; and
	 \item demonstration of breakdown of desired properties when sufficient conditions are not met using a classic data-driven null model.
\end{itemize}
In this way, we provide concrete guidelines for future methodological research into adjustment of indices using more general models by providing clear conditions under which favorable properties are guaranteed and highlighting the need for a more careful theoretical treatment when those conditions are not met. 

The paper is organized as follows. We formally introduce indices and discuss the general procedure for adjusting indices with respect to a general normalization and null model in Section \ref{sec:general}. Then we extend previous work on families of indices under adjustment, identifying sufficient conditions for favorable properties in Section \ref{sec:linear}. We provide two example index-model pairs in violation of the sufficient conditions which respectively are (a) non-positive after traditional adjustment for expectation and maximum rather than mean 0 and (b) identically 0 after statistical standardization. Section \ref{sec:disc} contains the discussion.

\section{General Adjustment for Chance}\label{sec:general}

Suppose we have two categorical variables we wish to compare where variable labels may or may not coincide. We might use an index to quantify similarity. Formally:

\begin{definition}[Index $S$]\label{defn:index}
Let $x$ and $y$ be two categorical variables observed for the same $k=1,...,N$ observations such that each $x_k$ takes one of $I$ possible values and each $y_k$ takes one of $J$ possible values. Define marginal counts $u,v$ with $u_i=\#\{k:x_k=i\},v_j=\#\{k:y_k=j\}$, and joint count $n$ with $n_{ij}=\#\{k:x_k=i,y_k=j\}$. An index $S$ for $x$ and $y$ is a real-valued function $S(n)$.
\end{definition}

Broadly, an index $S$ is a real function of contingency tables, that is joint count distributions $n$, sometimes for $n$ of a specific size (e.g., $I=J=2$ for comparing binary variables) and sometimes of any size (e.g., $I,J\in\mathbb N$ for comparing cluster labels) \citep{warrens2013association}. In practice, the index $S$ is desired to monotonically measure strength of similarity in some sense between $x$ and $y$ \citep{krippendorff1987}. We give two examples which give rise to many popular measures through adjustment:

\begin{definition}[Raw proportion $p$ of agreement]\label{defn:p_raw}
In the context of Definition \ref{defn:index}, suppose $x$ and $y$ take values in the same indexed set of $I=J$ possible values. The raw proportion $p$ of agreement between $x$ and $y$ is given by $p(n)=\sum_i\frac{n_{ii}}N.$
\end{definition}

\begin{definition}[Proportion $q$ of pair agreement]\label{defn:q_pair}
With the notation of Definition \ref{defn:index}, let $w$ be a count of a variable observed $N$ times taking values in a set $\mathcal R$ (e.g., $w=u$ and $\mathcal R=\{1,...,I\}$ or $w=n$ and $\mathcal R=\{1,...,I\}\times\{1,...,J\}$). The proportion $q$ of pairs of observations assigned the same value is given by $q(w)=\frac{\sum_{r\in\mathcal R}\binom{w_r}2}{\binom{N}2}\ .$
\end{definition}

While many similarity measures for continuous variables take expected value 0 in the presence of no association, this often is not the case for basic indices assessing shared information or association between categorical variables. In the cases of $p(n)$ and $q(n)$, values of 0 indicate observations generally unlikely to have resulted at random from independent variables $x$ and $y$. As such, these indices are typically adjusted through subtraction of an expected value and normalization by a maximum value to impose the properties that 0 indicates a level of observed association expected from unassociated variables and 1 indicates maximal association observable. In this way, more familiar association measures can be built, such as Cohen's $\kappa$ \citep{cohen1960coefficient}, Scott's $\pi$ \citep{scott1955reliability}, and Dice's coefficient \citep{dice1945measures} from $p(n)$ and the adjusted Rand index $ARI$ \citep{hubert1985comparing} and other pair-counting measures for clustering \citep{albatineh2006similarity} from $q(w)$.

\subsection{Adjustment of an Index}\label{sec:adj}

Index adjustment for expected and maximum values is traditionally linear in nature and requires specification of a normalization constant, typically a maximum level $S_{max}$ of association as measured by index $S$, and an expected value $\EE{S}$ with respect to a null model $\mathcal M$ \citep{scott1955reliability,cohen1960coefficient,fleiss1975measuring,hubert1985comparing,
krippendorff1987,albatineh2006similarity,warrens2008association}. Conceptually, we might expect the formulation
$$AS(n)=\frac{S(n)-\EE{S(\tilde n)}}{S_{max}-\EE{S(\tilde n)}},$$
where $\tilde n$ is distributed according to the null distribution, but this formulation is too rigid in practice. Sometimes we want to incorporate aspects of the observed count $n$ into our null or chance distribution, such as fixing observed marginal counts $u$ and $v$ when asserting a permutation model (Definition \ref{defn:Mperm}). Similar circumstances may arise in defining a maximum value, whereby the marginal counts $u$ and $v$ limit the maximum attainable value. As such, we must provide a definition for adjustment in very general terms. Unlike previous works, we choose to be fully explicit with respect to the null distribution and maximum and their potentially data-driven nature to help clarify unexpected behavior of index adjustment in a later section. We denote by $\amalg$ the disjoint union, that is the union of sets which naturally do not overlap.

\begin{definition}[Adjusted Index $AS$]\label{defn:adjindex}
Consider the context of Definition \ref{defn:index}. Let 
$$\mathcal M=\{M^{n'}:n'\in B\subseteq\amalg_{I,J}\mathbb N^{I\times J}\}$$ 
be a model for joint counts $\tilde n$, that is a collection of null distributions $M^{n'}$ on joint counts $\tilde n$ with distributions indexed by $n'\in B$, such that $\emph{\text{Support}}(n)\subseteq B$ and for each $n'\in \emph{\text{Support}}(n)$, $P_{M^{n'}}\left(\tilde n=n'\right)>0$ and $P_{M^{n'}}(\tilde n\in\emph{\text{Support}}(n))=1$. Let also $S_{max}^n$ be a real valued function of $n$. The adjusted index $AS(n)$ of $S(n)$ with respect to $\mathcal M$ and $S_{max}$ is given by
$$AS(n)=\frac{S(n)-\text{E}_{M^n}[S(\tilde n)]}{S_{max}^n-\text{E}_{M^n}[S(\tilde n)]}.$$
By convention, we take $AS(n)=0$ whenever $S_{max}^n-\text{E}_{M^n}[S(\tilde n)]=0$.
\end{definition}

Note, model indexing set $B$ contains the support of $n$ to ensure null distribution $M^n$ is always defined; $M^{n'}$ for $n'$ in the support of $n$ assigns total probability to the support of $n$ to ensure that impossible joint counts $\tilde n$ are not assigned positive probability under the null; and the null distribution $M^n$ always assigns positive probability to observed $n$. 

\subsection{Common Null Models and Normalizations}

Three standardly used data-driven models which respectively fix empirical marginal counts or fix theoretical marginal distributions based on unpooled or pooled empirical marginal counts are the permutation model $\mathcal M_{perm}$ (a hypergeometric model), the independence model for two populations $\mathcal M_{2,ind}$ (a multinomial model), and the independence model for one population $\mathcal M_{1,ind}$ (another multinomial model).
\begin{definition}[$\mathcal M_{perm}$]\label{defn:Mperm}
The permutation model $\mathcal M_{perm}=\{M^{n'}_{perm}:n'\in \amalg_{I,J}\mathbb N^{I\times J}\}$ is comprised of distributions $M^{n'}_{perm}$ arising as follows: let $x'$ and $y'$ be variables observed on $N$ observations with joint count $n'$ and marginals $u'$ and $v'$. Let $\pi$ be a permutation on $N$ observations selected uniformly at random. Then the joint count $\tilde n$ for observations $\tilde x_k=x_k'$ and $\tilde y_k=y_{\pi(k)}'$ has distribution $M^{n'}_{perm}$ given $u'$ and $v'$.
\end{definition}

\begin{definition}[$\mathcal M_{2,ind}$]\label{defn:Mind}
The independence model $\mathcal M_{2,ind}=\{M^{n'}_{2,ind}:n'\in \amalg_{I,J}\mathbb N^{I\times J}\}$ is comprised of distributions $M^{n'}_{2,ind}$ arising as follows: let $u'$ and $v'$ be marginal counts of joint count $n'$. Given $u'$ and $v'$, let $\tilde x$ and $\tilde y$ be variables with $N$ observations obtained as independent simple random samples from probability distributions $u'/N,v'/N$ respectively. Then the joint count $\tilde n$ for observations of $\tilde x$ and $\tilde y$ has distribution $M^{n'}_{ind}$ given $u'$ and $v'$.
\end{definition}

\begin{definition}[$\mathcal M_{1,ind}$]\label{defn:Mind1}
The independence model $\mathcal M_{1,ind}=\{M^{n'}_{1,ind}:n'\in \amalg_{I,J}\mathbb N^{I\times J}\}$ for one population is comprised of distributions $M^{n'}_{1,ind}$ arising as follows: let $u'$ and $v'$ be marginal counts of joint count $n'$. Given $u'$ and $v'$, let $\tilde x$ and $\tilde y$ be variables with $N$ observations each obtained as independent simple random samples from probability distribution $(u'+v')/(2N)$. Then the joint count $\tilde n$ for observations of $\tilde x$ and $\tilde y$ has distribution $M^{n'}_{1,ind}$ given $u'$ and $v'$.
\end{definition}

$S_{max}^n$ is standardly specified as an upper bound for $S(n)$, though this is not strictly necessary. When $S_{max}^n$ is properly obtained as a maximum, we express this maximum as $S_{max}^n=\max\limits_{n'\in\mathfrak X^n}S(n')$ for some sets of joint counts $\mathfrak X^n$ which may or may not depend on $n$. We consider a few natural choices of $\mathfrak X^n$, both previously considered \citep[see][]{cohen1960coefficient,warrens2013association}. 

We might maximize over the index domain, here understood as all possible joint counts on $N$ observations
$$\mathfrak X_D=\{n'\in \amalg_{I,J}\mathbb N^{I\times J}:\sum n'_{ij}=N\}.$$
Use of $\mathfrak X^n=\mathfrak X_D$ to define a \textit{domain maximum} $S^{n}_{D,max}$ will produce adjusted value 1 only when the index of the observation attains its global maximum value given the number of observations. Note that the domain set definition might be updated for context as needed, such as when marginals $u$ or $v$ are known a priori or the number of observations $N$ is not. We might also consider the support of the model 
$$\mathfrak X_{M^n}=\{n'\in\mathfrak X_D:P_{M^n}(\tilde n=n')>0\}.$$
Use of $\mathfrak X^n=\mathfrak X_{M^n}$ to define a \textit{model maximum} $S^n_{\mathcal M,max}$ will produce adjusted value 1 only when the observation is optimal under the support of the potentially data-driven null distribution $M^n$. The domain and model maxima both have historical usage \citep{davenport1991phi,cohen1960coefficient} and have been investigated for index adjustment \citep{warrens2013association}.

\subsection{\texorpdfstring{Impacts of the Choice of Generalized Normalization $S_{max}^n$}{Impacts of the Choice of Generalized Normalization Smax n}}\label{sec:max}

%\subsection{Impacts of the Choice of Generalized Normalization $S_{max}^n$}\label{sec:max}

Generally, for an adjusted index $AS$ to take values as high as 1 for some observable value, we need $S_{max}^n\leq S^{n}_{D,max}$, and to ensure an adjusted index $AS$ cannot take values greater than 1 for a joint count $\tilde n$ a priori observable according to the model $\mathcal M$, we need $S^n_{\mathcal M,max}\leq S_{max}^n$. Different choices of $S_{max}^n$ impose different conditions for which we consider an observed joint count $n$ to be optimal, thereby imposing different interpretations on resulting adjusted index $AS$. Consider the following examples from the literature:

When adjusting $p$ with respect to the permutation model, using the domain maximum produces adjusted index Cohen's $\kappa$ \citep{cohen1960coefficient}, which takes value 1 precisely when variables $x$ and $y$ are identical. Using the model maximum produces another adjusted index $\kappa/\kappa_m$ considered by \cite{cohen1960coefficient}, which takes value 1 precisely when as much agreement is observed as is attainable given differences in observed marginal counts $u$ and $v$. We can equivalently obtain Cohen's $\kappa$ through adjustment of $p$ under the independence model for two populations $\mathcal M_{2,ind}$ using the domain maximum. We obtain Scott's $\pi$ \citep{scott1955reliability,krippendorff1987} through adjustment of $p$ under the independence model for one population $\mathcal M_{1,ind}$ using the domain maximum, which also takes value 1 precisely when variables $x$ and $y$ are identical but assumes $x$ and $y$ are sampled from the same distribution rather than separate distributions as for $\mathcal M_{2,ind}$.

When adjusting $q(n)$ with respect to the permutation model, using the domain maximum produces an uninteresting index which takes value 1 precisely when $x$ and $y$ each assign all observations a single value. Instead, the standard maximum $q(n)\leq \frac12(q(u)+q(v))$, related to the domain maximum of the Rand index $RI=1-q(u)-q(v)+2q(n)$ \citep{rand1971objective}, produces the standard adjusted Rand index $ARI$ \citep{hubert1985comparing}. This adjustment takes value 1 precisely when $x$ and $y$ are identical up to relabelling. Taking a non-standard maximum $q(n)\leq\min(q(u),q(v))$, which \cite{warrens2022understanding} note corresponds with computing Loevinger's $H$ \citep{loevinger1947systematic,loevinger1948technic} from a pair contingency table (see Definition \ref{defn:pair}), results in a hierarchical adjusted Rand index which takes value 1 precisely when one of $x$ and $y$ can be deterministically obtained as a function of the other variable. In the common context of clustering where $x$ and $y$ are labels for two different groupings of the same observations, we understand a hierarchical adjusted Rand index to take value 1 precisely when the two groupings are related by hierarchy. We do not think the model maximum has been explicitly discussed for $q$ in the literature previously, and it appears less analytically tractable than other maxima. It would take value 1 precisely when as much pair agreement $q(n)$ as was possible given marginals $u$ and $v$ was observed.

Finally, we also note broader contexts fitting in this generalized notion of adjustment where $S^n_{max}$ is chosen to not be a maximum or upper bound for $S$. For example, it is traditional to fix $S^n_{max}=1$ for all indices \citep{albatineh2006similarity} while \cite{krippendorff1987} selects $S^n_{max}$ to be the value of $S$ under exhibition of maximal similarity by $n$ given contextual assumptions, which need not coincide with a maximum value of $S$. Furthermore, the above framework encapsulates statistical standardization of indices, as studied by \cite{romano2014standardized,romano2016adjusting}. For that purpose, we would take $S^n_{max}=\E{M^n}{S(\tilde n)}+\sigma_{M^n}(S(\tilde n))$, where $\sigma(\cdot)$ denotes a standard deviation.

\section{General Adjustment as an Operator}\label{sec:linear}

Towards investigation of favorable properties of the adjustment operator, such as idempotency, we next consider families of indices on which the operator may act in a similar fashion. In a certain sense, the generalized adjustment operator may be understood as linear in nature and so linear index families are natural to investigate. Linear families of indices which collapse under adjustment to a single adjusted index were first considered by \cite{albatineh2006similarity} for `pair-counting' indices derived from pair contingency tables and have been further considered in some generality by \cite{warrens2008association,warrens2008similarity,warrens2013association}, among others. We expand previous work for adjustment with respect to very general values $S_{max}^n$ and models $\mathcal M$. We then characterize contexts in which generalizations may fail.

\subsection{Favorable Properties Through Linear Families}

To contextualize the work in foundational literature \citep{albatineh2006similarity}, we begin with pair contingency tables. A pair contingency table assesses jointly in $x$ and $y$ the frequency with which a pair of observations is assigned the same or different $x$-values and respectively the same or different $y$-values. A pair of observations which is assigned the same value by $x$ and the same value by $y$ or assigned different values by $x$ and different values by $y$ would indicate similarity between $x$ and $y$ in how they partition observations. A pair of observations assigned the same value by one variable and different values by the other would indicate dissimilatiry between $x$ and $y$.
\begin{definition}[Pair contingency table]\label{defn:pair}
In the context of Definitions \ref{defn:index} and \ref{defn:q_pair}, the pair contingency table counting pairs of observations assigned the same or different values of $x$ and of $y$ is given below.
\begin{table}[h]
\renewcommand{\arraystretch}{1.5}
\centering
\scriptsize
\begin{tabular}{l||l|l}
 & Same $y$ & Different $y$\\ \hline \hline
Same $x$ & $q(n)\binom N2$ & $(q(u)-q(n))\binom N2$ \\ \hline
Different $x$ & $(q(v)-q(n))\binom N2$ & $(1-q(u)-q(v)+q(n))\binom N2$
\end{tabular}
\end{table}
\end{definition} 
\vspace{-3mm}
Indices designed to compare partitions of observations, such as those found through clustering, may depend on joint counts $n$ through the associated pair contingency table since it fosters direct comparison of partitions for which sets don't have meaningful or comparable labels as assigned by $x$ and $y$. These are typically referred to as \textit{pair-counting} indices, as opposed to information theoretic indices like Mutual Information and its variants \citep{shannon1948mathematical,vinh2009information}. Many pair-counting indices are adjusted using the permutation model which fixes marginal counts $u$ and $v$. Noting that the pair contingency table is a function of $n$ through $q(n)$, $q(u)$, and $q(v)$, this leads to the classic family of indices linear in $q(n)$ given $u$ and $v$.

\begin{definition}[$\mathcal L$; \cite{albatineh2006similarity}]
An index $S$ is said to be of the $\mathcal L$ family if, for some functions $\alpha$ and $\beta$ of $u$ and $v$, the index may be rewritten
$$S(n)=\alpha(u,v)+\beta(u,v)q(n)\ .$$
\end{definition}

This family $\mathcal L$ is specific to the index $q$ and to the decision to fix marginal counts $u$ and $v$. We define a broad collection of function families in a similar vein.

\begin{definition}[$\mathcal L_{\mathcal M}(S)$]\label{defn:family}
Consider the context of Definition \ref{defn:adjindex}. We define a linear family of indices $T$ with respect to the model $\mathcal M$ and index $S$:
\begin{align*}
\mathcal L_{\mathcal M}(S)=\left\{T:\forall n\ T(n)=\alpha_n+\beta_n S(n);\ \forall n'\in \mathfrak X_{M^n}\ \alpha_n=\alpha_{n'},\beta_n=\beta_{n'}\neq0\right\}\ .
\end{align*}
\end{definition}

Note that $\mathcal L_{\mathcal M_{perm}}(q)=\mathcal L$. For this general family, by demanding $\alpha_n$ and $\beta_n$ are constant on the supports of distributions $M^n$ in model $\mathcal M$, we can relate expectations of indices $T\in\mathcal L_{\mathcal M}(S)$ and $S$. During adjustment, we also choose to relate bounds $T_{max}^n$ and $S_{max}^n$ through linearity, though notions of a maximum may be violated by doing so
\begin{align}
	\text{E}_{M^n}[T(\tilde n)] & =\alpha_n+\beta_n\text{E}_{M^n}[S(\tilde n)]\nonumber\\
	T_{max}^n & =\alpha_n+\beta_n S_{max}^n\ .\label{eqn:pmax}
\end{align}

We give straightforward variations on Theorem 9 and Lemmas 10,12 of work by  \cite{warrens2013association} in this more general context.
\begin{theorem}\label{thm:L}
%Consider the context of Definition \ref{defn:adjindex}. 
Given $T\in\mathcal L_{\mathcal M}(S)$ and $S_{max}$, define $T_{max}$ according to Equation \eqref{eqn:pmax}. Then the adjustment $AT(n)$ of $T(n)$ with respect to $\mathcal M$ and $T_{max}$ is equivalent to the adjustment $AS(n)$ of $S(n)$ with respect to $\mathcal M$ and $S_{max}$: $AT(n)=AS(n)$.
\end{theorem}
\begin{corollary}\label{cor:idem} Consider the context of Theorem \ref{thm:L}. If $S_{max}^n$ and $\E{M^n}{S(\tilde n)}$ are each constant in $n$ on the support of each $M^{n'}$, then $\E{M^n}{AS(\tilde n)}\equiv 0$, $\mathcal L_{\mathcal M}(S)$ is closed under $A$, and, defining $AS_{max}$ according to Equation \eqref{eqn:pmax}, $A$ is idempotent ($A^2=A$) on $\mathcal L_{\mathcal M}(S)$. 
\end{corollary}
%\begin{corollary}\label{cor:perm} Consider the context of Theorem \ref{thm:L} for $\mathcal M=\mathcal M_{perm}$. If $S_{max}^n$ is constant in $n$ on the support of each $\mathcal M_{perm}^{n'}$, then $\mathcal L_{\mathcal M_{perm}}(S)$ is closed under $A$ and, defining $AS_{max}$ according to Equation \eqref{eqn:pmax}, $A$ is idempotent ($A^2=A$) on $\mathcal L_{\mathcal M_{perm}}(S)$. 
%\end{corollary}
Theorem \ref{thm:L} and Corollary \ref{cor:idem} can be reframed for statistical standardization:
\begin{corollary}\label{cor:standard}
%Consider the context of Definition \ref{defn:family}. 
Let $A$ denote the statistical standardization operator on $\mathcal L_{\mathcal M}(S)$. That is, if $\tilde n\sim M^n$ has null distribution given $n$, $A$ adjusts indices $T\in\mathcal L_{\mathcal M}(S)$ with respect to $T_{max}^n=\E{M^n}{T(\tilde n)}+\sigma_{M^n}(T(\tilde n))$ and $\mathcal M$. Then for all $T=\alpha_n+\beta_n S\in\mathcal L_{\mathcal M}(S)$, $AT(n)=\text{\emph{sgn}}(\beta_n)AS(n)$. If $\E{M^n}{S(\tilde n)}$ and $\sigma_{M^n}(S(\tilde n))>0$ are also each constant in $n$ on the support of each $\mathcal M^{n'}$, $AT(\tilde n)$ has mean 0 and variance 1 with respect to $M^n$.
\end{corollary}

The treatment of the $\mathcal L_{\mathcal M}(S)$ families described in this section differs significantly from the original treatment of $\mathcal L$ by \cite{albatineh2006similarity} in that, for $T\in\mathcal L_{\mathcal M}(S)$, we enforce that $T_{max}^n$ is the same linear function of $S_{max}^n$ that $T$ is of $S$ (Equation \ref{eqn:pmax}). Work by \cite{albatineh2006similarity} with the family $\mathcal L$ instead fixes $T_{max}^n=1$ regardless of coefficients $\alpha(u,v)=\alpha_n$ and $\beta(u,v)=\beta_n$; note that that treatment requires a fully general understanding of $S_{max}^n$ as potentially not even an upper bound. As observed in that work and others, most pair-counting indices of interest in fact take domain maximum of 1. However, this unnatural fixing of $T^n_{max}=1$ for more general consideration of linear transformations of an index results in infinitely many subfamilies parametrized by the coefficient ratio $\frac{1-\alpha}\beta$ of indices equivalent after adjustment. It also fails to acknowledge the richness of available indices and their interpretations that may be obtained through selection of other maxima. Though the convention of fixed $T_{max}^n=1$ is standard across the literature \citep{warrens2008similarity,warrens2022understanding,krippendorff1987,
fleiss1975measuring}, we adopted the convention of variable $T_{max}^n$ \citep{warrens2013association,warrens2008association} as more natural for linear transformations.

\subsection{Violations of Intended Properties}

Corollaries \ref{cor:idem} and \ref{cor:standard} establish the basic desired properties of adjustment under sufficient conditions: 
\begin{itemize}[noitemsep,topsep=0pt,label=--]
	\item adjusted measures have mean 0 under the null,
	\item statistically standardized measures also have variance 1 under the null, and 
	\item adjustment is idempotent, meaning adjusted measures do not change under further adjustment with respect to the same model and related maximum (per Equation \ref{eqn:pmax}).
\end{itemize}
The sufficient conditions are constancy conditions, namely that for adjustment $A$ with respect to $\mathcal M$ and $S_{max}$, $S_{max}^n$ and $\E{M^n}{S(\tilde n)}$ are each constant in $n$ on the support of each $M^{n'}$. This says that neither upper bounds/normalizing constants nor expectations can be random quantities with respect to the null distribution. For domain and model maxima $S_{max}$, this is intuitive and constancy conditions are always satisfied. For expectations and standard deviations with respect to null model $\mathcal M$, this requirement is a little more subtle because of the null distribution being potentially data-driven. We focus on expectations with standard deviations being analogous.

The condition that $\text{E}_{M^n}[S(\tilde n)]$ is constant on the support of each $\mathcal M^{n'}$ is a sufficient condition which is satisfied for all indices $S$ when either distributions $M^n=M$ do not depend on observations $n$ or when distributions $M^n$ depend on $n$ through a property $r(n)$ for which $r(\tilde n),\ \tilde n\sim M^n$ is constant in $\tilde n$: $r(\tilde n)\equiv r(n)$. An example of the first option is uniform distributions over clustering ensembles \citep{gates2017impact}. The second option requires that a data-driven null distribution deterministically fix whatever observed properties it incorporates, e.g., fixing observed marginal distributions as in the permutation distribution when marginal distributions are a priori unknown.

\textit{Thus, constancy conditions are broadly met by null models which are not data-driven or which depend only on properties of observed data which are then not subjected to randomness by the null. Desired properties of adjustment follow.}

This renders the historical use of the permutation model special in that the permutation model is data-driven but Corollary \ref{cor:idem} still applies. Theoretical work done with permutation models need not generalize cleanly to new data-driven null models. Generally, the nested expectations $\text{E}_{M^n}[\text{E}_{M^{\tilde n}}[S(\tilde{\tilde n})]]$ considered when investigating properties of adjusted indices rely on two different null distributions, $\tilde n\sim M^n$ based on observed $n$, and $\tilde{\tilde n}\sim M^{\tilde n}$ based on $\tilde n$. These nested expectations naturally collapse under constancy conditions, leading to basic desired properties, but need not otherwise. In particular, the equivalence $\E{M^n}{S(\tilde n)}\equiv\E{M^n}{\E{M^{\tilde n}}{S(\tilde{\tilde n})}}$ for $\tilde n|n\sim M^n$ and $\tilde{\tilde n}|\tilde n\sim M^{\tilde n}$ should not be assumed without explicit verification for a data-driven model $\mathcal M$ and index $S$.

\textit{When a data-driven null model subjects observed distributional properties to randomness, desired properties of adjustment need not hold and must be otherwise confirmed.}

Indeed, we find violations of the results of Corollary \ref{cor:idem} when constancy conditions are not met. As proof by examples, we consider an index which is non-positive after traditional adjustment and an index which is deterministically, identically 0 after statistical standardization.
\begin{proposition}\label{prop:fail} Consider the context of Theorem \ref{thm:L}. For each of the following properties, there exist triples $(S,S_{max},\mathcal M)$ such that $\text{E}_{M^n}[S(\tilde n)]$ is not constant in $n$ on the support of each $\mathcal M^{n'}$ and which violate the property:
\begin{enumerate}[label={(\arabic*.)}]
	\item for $\tilde n|n\sim M^n$ and $\tilde{\tilde n}|\tilde n\sim M^{\tilde n}$, $\E{M^n}{S(\tilde n)}\equiv\E{M^n}{\E{M^{\tilde n}}{S(\tilde{\tilde n})}}$,  
	\item $AS\in\mathcal L_{\mathcal M}(S)$,
	\item $\E{M^n}{AS(\tilde n)}\equiv0$, and
	\item if $S_{max}$ and $AS_{max}$ are both domain maxima, $A^2S(n)\equiv AS(n)$.
\end{enumerate}
If additionally $S_{max}^n=\E{M^n}{S(\tilde n)}+\sigma_{M^n}(S(\tilde n))$ and $\sigma_{M^n}(S(\tilde n))>0$ for some $n$, the following property can be violated:
\begin{enumerate}
\item[(5.)] $AS(\tilde n)$ has mean 0 and variance 1 under the null distribution $\tilde n\sim M^n$.
\end{enumerate}
%(Proof by examples in Section \ref{sec:proof} using $\mathcal M_{2,ind}$.)
\end{proposition}
\textit{Remark 1:}  %See Section \ref{sec:corrections} for a discussion of how Proposition \ref{prop:fail} compares to the literature.
Violation of closure under adjustment (2.) precludes the possibility of idempotency as stated in Corollary \ref{cor:idem}. Specifically, the Corollary uses $AS\in\mathcal L_{\mathcal M}(S)$ to relate $S_{max}$ and $AS_{max}$ according to Equation \eqref{eqn:pmax} in defining $A^2$. For completeness, we investigate using domain maxima for both $S_{max}$ and $AS_{max}$, and find the modified notion of idempotency (4.) can still fail. We note, however, that use of domain maxima in place of Equation \eqref{eqn:pmax} is not necessarily natural and can lead to breakdown of Theorem \ref{thm:L}, as already discussed for adjustments of $q(n)$ and the Rand index $RI$ with respect to the permutation distribution and domain maximum (Section \ref{sec:max}).

%\subsection{Proof by examples of Proposition \ref{prop:fail}}\label{sec:proof}
\subsection{Examples of Violation}\label{sec:proof}

\begin{proof}of Proposition \ref{prop:fail}.
First, we provide an example index $S$, domain maximum $S_{max}$, and model $\mathcal M$ for which properties 1-4 are violated. Note that violation of property 1 implies the non-constancy assumption of the Proposition, and so non-constancy is not separately addressed. For a more tractable example, we consider the toy index $S(n)=u_1^2$ with the independence model $\mathcal M_{2,ind}$ for two populations and domain maximum $N^2$.

\textbf{(1.)} Note that if $\tilde n\sim M^n_{2,ind}$, then $\tilde u_1\sim$ Binom$(N,u_1/N)$. Nested expectations do not naturally simplify due to dependency of $M_{2,ind}^n$ on $n$, that is non-constancy of $\E{M_{2,ind}^n}{\tilde u_1^2}$ in $n$ over supports of each $M_{2,ind}^{n'}$:
\begin{align*}
\E{M_{2,ind}^n}{\tilde u_1} & =N\frac{u_1}N=u_1
\\
\sigma^2_{M_{2,ind}^n}\left(\tilde u_1\right) & =N\frac{u_1}N\left(1-\frac{u_1}N\right)=u_1\left(1-\frac{u_1}N\right)
\\
\E{M_{2,ind}^n}{\tilde u_1^2} & 
= u_1\left(1-\frac{u_1}N\right)+u_1^2
= u_1+\frac{N-1}Nu_1^2
\\
\E{M_{2,ind}^n}{\E{M_{2,ind}^{\tilde n}}{\tilde{\tilde u}_1^2}} & 
= \E{M_{2,ind}^n}{\tilde u_1+\frac{N-1}N\tilde u_1^2} 
\\
 & = \frac{2N-1}Nu_1+\left(\frac{N-1}N\right)^2u_1^2
\neq \E{M_{2,ind}^n}{\tilde u_1^2}\ .
\end{align*}

\textbf{(2.)} We compute the adjusted index when $S_{max}\neq\E{M_{2,ind}^n}{S(\tilde n)}$, that is when $u_1\neq N$:
\begin{align*}
AS(n) & =\frac{u_1^2-u_1-\frac{N-1}Nu_1^2}{N^2-u_1-\frac{N-1}Nu_1^2}=\frac{u_1-N}{u_1-N}\cdot\frac{-u_1}{N^2+(N-1)u_1}\ .
\end{align*} We see $AS(n)$ is not linear in $u_1^2$, specifically $AS\notin\mathcal L_{\mathcal M_{2,ind}}(S)$, as the rational function $\frac{-u_1}{N^2+(N-1)u_1}$ cannot be formulated as a quadratic $\alpha+\beta u_1^2$ for constants $\alpha,\beta$ and integer $0<u_1<N$ for large $N$. 

\textbf{(3.)} By convention, $AS(n)=0$ when $S_{max}=\E{M_{2,ind}^n}{S(\tilde n)}$, that is when $u_1=N$. Thus, $AS(n)\leq 0$ with $AS(n)=0$ precisely when $u_1=0,N$, and so $\E{M_{2,ind}^n}{AS(\tilde n)}<0$ for $0<u_1<N$. 

\textbf{(4.)} By reasoning from property (3.), we have domain maximum $AS_{max}=0$. 

The expected value $\E{M_{2,ind}^n}{AS(\tilde n)}$ of the adjusted index and the value of the doubly adjusted index $A^2S(n)$ can be computed
\begin{align*}
A^2S(n) & =\frac{AS(n)-\E{M_{2,ind}^n}{AS(\tilde n)}}{0-\E{M_{2,ind}^n}{AS(\tilde n)}}\\
\E{M_{2,ind}^n}{AS(\tilde n)} & =\sum_{k=1}^{N-1}\frac{-k}{N^2+(N-1)k}\binom Nk\left(\frac{u_1}N\right)^k\left(1-\frac{u_1}N\right)^{N-k}\ .
\end{align*}
We can disprove idempotency by making a selection of $u_1$ and $N$ and computing, e.g., $u_1=1$ and $N=2$ gives $AS(n)=-\frac15\neq-1=A^2S(n)$. Perhaps more satisfactorily, we provide plots for integer $1\leq u_1<N\leq 100$ to show the generally non-zero nature of the difference; see Figure \ref{fig:adj_u2}, middle column, in Appendix \ref{sec:proof_c}.

\textbf{(5.)} For this property, we consider a different toy index. Let $S(n)=u_1$, $\mathcal M=\mathcal M_{2,ind}$, $\tilde n\sim M^n_{2,ind}$, and $0<u_1<N$. As before, $\tilde u_1\sim$ Binom$(N,u_1/N)$ with mean $u_1$ and variance $u_1(1-u_1/N)>0$. For $0<u_1<N$, we have
\begin{align*}
AS(n) & =\frac{u_1-u_1}{\sqrt{u_1\left(1-\frac{u_1}N\right)}}\equiv 0\ .
\end{align*}
As per convention, $AS(n)=0$ also when $u_1=0,N$. Thus, under the null distribution $\tilde n\sim M^n_{2,ind}$, the standardized index $AS(\tilde n)$ is identically 0 and does not have variance 1.
\end{proof}

\noindent\textit{Remark 2:} These proofs by example of non-zero expectation (3.), lack of idempotency (4.), and failed statistical standardization (5.) may appear fundamentally dependent on the convention $AS(n)=0$ for $S^n_{max}=\text{E}_{M^n_{2,ind}}[S(\tilde n)]$. This convention is chosen in Definition \ref{defn:adjindex} because it indicates no similarity observed beyond expectation ($AS(n)=0$) when the expectation is maximal possible similarity ($S^n_{max}=\text{E}_{M^n_{2,ind}}[S(\tilde n)]$). If we choose a different convention, $AS(n)=c$ for $S^n_{max}=\text{E}_{M^n_{2,ind}}[S(\tilde n)]$, the results still hold; see Appendix \ref{sec:proof_c}. Note that because the independence model $\mathcal M_{2,ind}$ has some distributions $M^n_{2,ind}$ which assign positive probability to every joint count in the domain $\mathfrak X_D$ and others which assign total probability to a single joint count, such a convention must be chosen in order for nested expectations and double adjustment to be well-defined.

\section{Conclusion}\label{sec:disc}

We investigated theoretical properties of index adjustment with respect to fully general models and normalization constants, establishing sufficient conditions for favorable properties, namely for general index adjustment to be idempotent and result in mean 0 adjusted indices. 

We further found that adjustment with respect to some models can produce adjusted indices in violation of the defining desired properties. These violations arise when distributions used in adjustment are data-driven. Permutation distributions, i.e., hypergeometric distributions, appear as special because while they are typically data-driven, they also fix observed properties and meet sufficient conditions for favorable adjustment properties as a result. This frames the foundational work \citep{albatineh2006similarity} as specific in its focus on the permutation distribution and further motivates modern \citep{gates2017impact} and classic \citep{krippendorff1987} interest in the choice of models besides the permutation model.

Future theoretical and empirical work in this area will benefit from more precise attention to models used in adjustment. Where possible, consideration of models, indices, and normalizations which meet the constancy assumptions of Corollary \ref{cor:idem}, as by \cite{gates2017impact}, would ensure favorable properties.

\subsection{Future Work}

The treatment of index adjustment in this work presumes that favorable properties should hold exactly under the null distribution. In practice, it may be sufficient if these properties hold approximately. For example, Scott's $\pi$, which is the adjustment of $p$ under the data-driven independence model for 1 population with respect to a domain maximum, fails to satisfy constancy conditions and does not have mean 0 under the null distribution. \cite{scott1955reliability} provided a large sample characterization instead. Future work might investigate sufficient conditions for favorable asymptotic behavior when constancy conditions are not met. Clustering will supply a particularly interesting context for such investigation as not only may the number of labels vary under the null distribution, but the maximum possible number of labels may grow unbounded with sample size. Extensions to fuzzy clustering with different null models \citep{dewolfe2025random} might also be considered.

\appendix
%\begin{appendix}
%\section{Appendix: Proposition \ref{prop:fail} parts (3.), (4.), and (5.) allowing different conventions $c$ for when the maximal index value is expected.}\label{sec:proof_c}
\section{\texorpdfstring{Proposition \ref{prop:fail} parts (3.), (4.), and (5.) allowing different conventions $c$ for when the maximal index value is expected.}{Proposition \ref{prop:fail} parts (3.), (4.), and (5.) allowing different conventions c for when the maximal index value is expected.}}\label{sec:proof_c}

\FloatBarrier

In the context of the proof of Proposition \ref{prop:fail}, violation of properties (3.), (4.), and (5.), we consider the newly relaxed convention that when $\E{M^n_{2,ind}}{S(\tilde n)}=S^n_{max}$, we assign fixed value $AS(n)=c$. For reference, two a priori natural choices of $c$ would be $c=0$, as originally considered, suggesting no similarity beyond expectation when maximal similarity is expected, and $c=1$, suggesting perfect similarity beyond expectation when maximal similarity is expected as well as ensuring domain maximum $AS_{max}=1$. 

\begin{figure}[!ht]
%\vspace{-25mm}
\centering
\includegraphics[width=\textwidth, trim={0 40mm 0 35mm},clip]{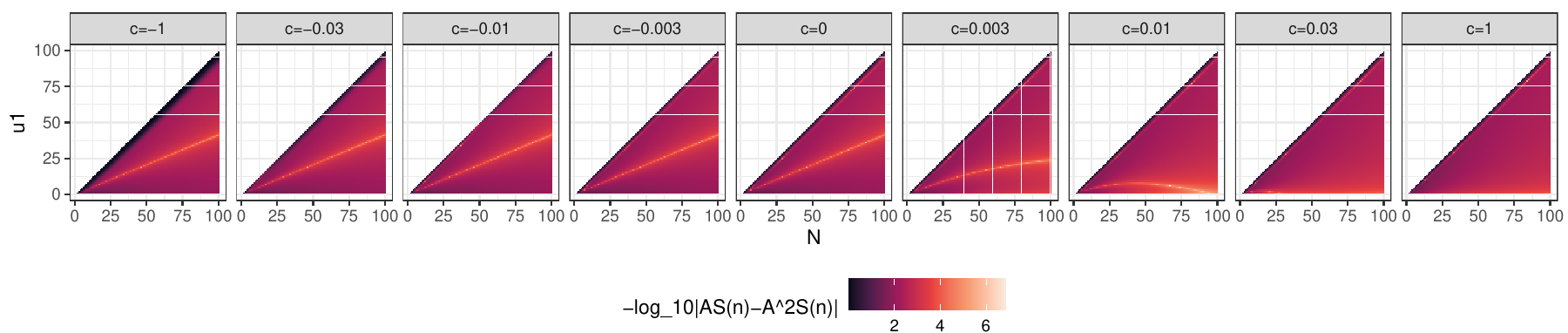}
%\vspace{-10mm}
\caption{\label{fig:adj_u2} Visualization of violation of idempotency, i.e., visualization of the general non-equivalence of $AS$ and $A^2S$, for an example index $S$. Specifically, for $S(n)=u_1^2$, $S_{max}=N^2$, $AS_{max}=\max(0,c)$, we have heatmaps of $-\log_{10}|AS(n)-A^2S(n)|$ as a function of $u_1$, $N$, and convention $c$ for the adjusted value of an index $S$ when $S_{max}=\E{M^n}{S(\tilde n)}$. Larger log differences indicate approaching equivalence.}
%\vspace{-10mm}
\end{figure}

\begin{proof}
\textbf{(3.)} Recall $AS(n)=c$ for $u_1=N$ and, for $u_1<N$,
\begin{align*}
AS(n) & =\frac{-u_1}{N^2+(N-1)u_1}\ .
\end{align*}
We compute the expected value $\E{M^n_{2,ind}}{AS(\tilde n)}$ of $AS$ when $0<u_1<N$, noting for $\tilde n\sim M_{2,ind}^n$, $\tilde u_1\sim$ Binom$(N,u_1/N)$
\begin{align*}
\E{M^n_{2,ind}}{AS(\tilde n)} & =c\left(\frac{u_1}N\right)^N +\sum_{k=0}^{N-1}\frac{-k}{N^2+(N-1)k}\binom Nk\left(\frac{u_1}N\right)^k\left(1-\frac{u_1}N\right)^{N-k}\ .
\end{align*}
Assume temporarily that $\E{M^n_{2,ind}}{AS(\tilde n)}\equiv0$. Then we can solve for $c$ for $0<u_1<N$
\begin{align*}
c & =\sum_{k=1}^{N-1}\frac{k}{N^2+(N-1)k}\binom Nk\left(\frac N{u_1}-1\right)^{N-k}\ .
\end{align*}
As $c$ is constant, not a strictly decreasing function of $u_1$, it cannot be that $\text{E}_{M^n_{2,ind}}[AS(\tilde n)]\equiv0$.

\textbf{(4.)} To investigate idempotency, we observe $\max(0,c)$ to be the domain maximum for $AS(n)$
\begin{align*}
A^2S(n) & =\frac{AS(n)-\text{E}_{M^n_{ind}}[AS(\tilde n)]}{\max(0,c)-\text{E}_{M^n_{ind}}[AS(\tilde n)]}\ .
\end{align*}
As $c=0$ is handled in Section \ref{sec:proof}, we restrict considerations to $c\neq0$. 

Let $f$, $p$, and $g$ respectively denote the function of $u_1$ corresponding with $AS$ when $u_1<N$, the binomial probabilities for $\tilde u_1$, and the expected value of $AS(\tilde n)(1-\delta_{\tilde u_1}^N)$ where $\delta_x^y$ is 1 when $x=y$ and 0 otherwise.
\begin{align*}
f(u_1,N) & =\frac{-u_1}{N^2+(N-1)u_1}
& f_{u_1}(u_1,N) & =\frac{-N^2}{(N^2+(N-1)u_1)^2}<0
\\
p(u_1,k,N) & =\binom Nk\left(\frac{u_1}N\right)^k\left(1-\frac{u_1}N\right)^{N-k}
& 1 & =\sum_{k=0}^Np(u_1,k,N)
\\
g(u_1,N) & =\sum_{k=0}^{N-1}f(k,N)p(u_1,k,N)
 & \E{M_{2,ind}^n}{AS(\tilde n)} & =c\left(\frac{u_1}N\right)^N+g(u_1,N)
\end{align*}
Note 
$$\frac{-1}{N}<f(N-1,N)(1-p(u_1,N,N))\leq g(u_1,N)\leq f(1,N)(1-p(u_1,N,N))<\frac{-1}{4N^2},$$ 
and so $g(u_1,N)$ converges uniformly to 0 as $N\rightarrow\infty$ for $0<u_1<N$. Suppose $u_1=N-j$ for fixed $j$ and note $AS(n)=f(N-j,N)<0$. Then, denoting the indicator function by $\mathbbm 1(\cdot)$, we have
\begin{align*}
\frac1N\frac{A^2S(n)}{AS(n)} & =\frac{1}{N}\left[1-\frac{c\left(\frac{N-j}N\right)^N+g(N-j,N)}{f(N-j,N)}\right]*\\
 & \hspace{15mm}\left(\max(0,c)-c\left(\frac{N-j}N\right)^N-g(N-j,N)\right)^{-1}\\
 & =\left[\frac{1}{N}+\left(c\left(\frac{N-j}N\right)^N+g(N-j,N)\right)\frac{N^2+(N-1)(N-j)}{N(N-j)}\right]*\\
 & \hspace{15mm}\left(\max(0,c)-c\left(\frac{N-j}N\right)^N-g(N-j,N)\right)^{-1}\\
 & \xrightarrow{N\rightarrow\infty}\left[0+(ce^{-j}-0)2\right]\left(\max(0,c)-ce^{-j}-0\right)^{-1}=\frac{2}{e^j\mathbbm 1(c>0)-1}\neq 0\ .
\end{align*}
We conclude for $N\gg j\geq1,u_1=N-j$ that $A^2S(n)\approx\frac{-1}{e^j\mathbbm 1(c>0)-1}$, $AS(n)\approx\frac{-1}{2N}$. Thus, $A^2(n)\not\equiv AS(n)$.

For reference, we provide plots of $AS(n)-A^2S(n)$ for integer $1\leq u_1<N\leq 100$ and various choices of $c$ (Figure \ref{fig:adj_u2}). 

\textbf{(5.)} Recall that for this property, we consider a different toy index. Let $S(n)=u_1$ and $\mathcal M=\mathcal M_{2,ind}$, and recall for $\tilde n\sim M^n_{2,ind}$, we have $\tilde u_1\sim$ Binom$(N,u_1/N)$. For $0<\tilde u_1<N$, we have
\begin{align*}
AS(n) & =\frac{\tilde u_1-\tilde u_1}{\sqrt{\tilde u_1\left(1-\frac{\tilde u_1}N\right)}}\equiv 0\ .
\end{align*}
As per convention, $AS(\tilde n)=c$ when $\tilde u_1=0,N$. Thus, we have
\begin{align*}
AS(\tilde n) & =c(\delta_{\tilde u_1}^0+\delta_{\tilde u_1}^N)\\
\E{M^n_{2,ind}}{AS(\tilde n)} & =c P_{M^n_{2,ind}}(\tilde u_1=0,N)=c\left(\left(\frac{u_1}N\right)^N+\left(1-\frac{u_1}N\right)^N\right)\ .
\end{align*}
As $\left(\frac{u_1}N\right)^N+\left(1-\frac{u_1}N\right)^N>0$ for all $u_1$, we have $\E{M^n_{2,ind}}{AS(\tilde n)}\equiv 0$ only if $c=0$. If $c=0$, $AS(\tilde n)$ is identically 0 and has 0 variance.
\end{proof}

\FloatBarrier
%\end{appendix}

\subsubsection*{Acknowledgements}

This work was supported by the National Institutes of Health under Grants R01 HL142049 and T32 HL007085.

\subsubsection*{Conflicts of Interest}

Authors have no conflicts to declare.

\bibliography{references}

\end{document}